# Chemical Vapor Deposition-Assembled Graphene Field-Effect Transistor on Hexagonal Boron Nitride

Edwin Kim, Tianhua Yu, Eui Sang Song, and Bin Yu*

College of Nanoscale Science and Engineering, State University of New York, Albany, NY 12203

**ABSTRACT**  We investigate key electrical properties of monolayer graphene assembled by chemical-vapor-deposition (CVD) as impacted by supporting substrate material. Graphene field-effect transistors (GFETs) were fabricated with carbon channel placing directly on hexagonal boron nitride (h-BN) and $SiO_2$, respectively. Small-signal transconductance ($g_m$) and effective carrier mobility ($\mu_{eff}$) are improved by 8.5 and 4 times on h-BN, respectively, as compared with that on $SiO_2$. Compared with GFET with exfoliated graphene on $SiO_2$, $g_m$ and $\mu_{eff}$ measured from device with CVD graphene on h-BN substrate exhibits comparable values. The experiment demonstrates the potential of employing h-BN as a platform material for large-area carbon electronics.

* Corresponding Author.  byu@uamail.albany.edu

Graphene, a two-dimensional (2D) carbon sheet arranged in hexagonal honeycomb structure,[1] is an excellent material candidate for a spectrum of applications in nanoelectronics. Since its discovery, a number of attractive properties have been reported, including ultra-high carrier mobility[2,3,4] and ambipolar conduction,[5,6] making graphene potentially useful for functional nano-device implementation. Superb carrier mobility translates into fast speed for information processing. Graphene is also photoconductive in the IR/near IR spectral range.[7] To date, the majority of experimental devices have been reported with mechanically exfoliated graphene sheets. Although epitaxial growth of graphene on silicon carbide (SiC) substrate has been reported,[8] chemical vapor deposition (CVD), employing metal (e.g., nickel or copper) as growth catalyst, is considered as a potential route towards large-scale, on-chip direct assembly of graphene.[9,10,11,12]

The substrate material is critical to carrier transport in graphene, a 2D system sensitive to charged impurity or defects. Silicon dioxide ($SiO_2$) with selected thickness is typically used as the substrate.[13] Recently, hexagonal boron nitride (h-BN) was proposed as an alternative supporting material.[14] h-BN is an insulating isomorph of graphite with B and N atoms occupying the equivalent *A* and *B* sub-lattices in the *Bernal* configuration. It is relatively inert and expected to be free of dangling bonds or surface charge traps. The atomically flat surface helps to suppress surface roughness and charge density inhomogeneity as well as rippling in graphene.[15] With the bandgap energy of $E_G = 5.97$ eV and dielectric constant of $\varepsilon \approx 4$ (close to that of $SiO_2$), h-BN has small lattice mismatch (1.7%) with graphene. Excellent carrier mobility in graphene was reported on h-BN substrate (~10 times higher than that on $SiO_2$).

In this work, an ultra-thin h-BN multilayer, placed underneath the CVD-assembled graphene sheet, serves as the supporting substrate. The device fabrication process is schematically shown in Fig. 1a. A-few-layer h-BN flakes were mechanically exfoliated from the synthesized h-BN nanocrystal[16] onto 100 nm $SiO_2$-coated Si substrate (Boron-doped). A monolayer graphene was then transferred onto h-BN sheet after Cu-catalytic CVD growth. Graphene growth began using $CH_4$ (300 Sccm) as precursor for 30 minutes after Cu pre-anneal process at 1000ºC. Once graphene monolayer was grown on Cu surface, polymethyl-methacrylate (PMMA) was spin-coated on one side of graphene and Cu was etched away using wet chemistry ($FeCl_3·6(H_2O)$ solution). After the PMMA (with graphene) was transferred onto the target substrate, it was removed with acetone and graphene monolayer positions itself on the target substrate where the location of h-BN flakes is identified with optical microscope.

Raman spectra were taken to examine monolayer graphene and h-BN. In Fig. 1b, position "0" is the transferred graphene. Its Raman spectrum shows the two peaks in G- and 2D-lines.[17] A thin h-BN flake was placed underneath the graphene at position "1", presenting a weak peak in D-line at about 1366 $cm^{-1}$, a signature of h-BN.[18] Thus h-BN was well identified along with the transferred monolayer graphene. As h-BN gets thicker at position "2", the peak in D-line increases while the peaks in G- and 2D-lines decrease (The G/2D ratio maintains roughly unchanged). The modulation of signature peaks in Raman spectra may be due to the composition of h-BN and graphene in the depth of focus.[19, 20] The Raman photons are collected from where an object is focused. As the h-BN film underneath the CVD-assembled graphene becomes thicker, more photons are obtained from the increased h-BN layer within the illuminated volume.

Once h-BN and graphene were locally identified, a graphene-etching step was involved to isolate the device. Positive photoresist (PMMA) was spin-coated, followed by electron-beam lithography (EBL) process. After pattern development in 1:3 MIBK:IPA, the graphene active area covered by PMMA was protected, while the uncovered area was removed during the plasma etching (performed at 0.1 Torr, 10 sccm of $O_2$, and 27 Watts of RF power at 30 KHz). The source and drain metal contacts of the graphene field-effect transistor (GFET) were made by EBL patterning, Ti (5 nm)/Au (30 nm) evaporation, and a subsequent lift-off process.

To investigate the substrate impact on graphene electrical behavior, we made two GFETs on $SiO_2$ and h-BN substrate, respectively (Fig. 2a). Each device has three distinct gate lengths ($L_G$'s): 2.2 µm, 3 µm, and 5 µm. 500 nm-wide graphene stripes were patterned with plasma etching and three adjacent transistors were fabricated with a common source. Fig. 2b shows the graphene channel resistance ($R$) as a function of back gate bias ($V_{BG}$). The $R$-$V_{BG}$ curve indicates that transferred CVD-grown graphene on $SiO_2$ substrate is heavily p-type doped as previously reported.[21] On the other hand, graphene on h-BN substrate only shows moderate n-type doping with much narrower peaks at the Dirac point. The p-type doping for graphene on $SiO_2$ is possibly due to water molecules.[22, 23] Silanol groups (SiOH), with OH coupled to the dangling bond at $SiO_2$ surface, is highly hydrophilic.[24] As hexamethyldisilazane (HMDS) coated on $SiO_2$ screens graphene from the OH-terminated hydrophilic substrate,[25] the h-BN layer protects graphene from substrate deficiency. The weaker n-type doping of graphene on h-BN is assumed to be caused by the PMMA residue.[26]

The total drain-to-source resistance is represented by $R_T = 2R_m + 2R_c + R_g(V_G)$ where $R_m$, $R_c$, and $R_g$ are the resistance of metal, metal/graphene contact, and graphene, respectively. $V_G$ is the back gate voltage. The graphene resistance is also given by $R_g(V_G) = R_{sh}(V_G)L/W$ where $R_{sh}(V_G)$ is the sheet resistance. $L$ and $W$ are the channel length and width of the GFET, respectively. The contact resistance is extracted by $R_c = \frac{R_2(V_G)L_1 - R_1(V_G)L_2}{2(L_1 - L_2)}$ in the multi-contact configuration in which "1" and "2" denote each of the two terminals. With reasonable accuracy, the contact resistances are calculated to be ~2 KΩ for devices *1-2* and *1-3* on $SiO_2$ and ~3 KΩ for devices *1-2* and *1-4* on h-BN. Here the pair of number represents the location of source-drain contact for a GFET. Ignoring metal resistance, the difference of $R_T$ for a pair of two devices, i.e., between *1-2* and *1-3* and between *1-2* and *1-4*, is primarily caused by the difference in gate length.

Small-signal transconductance ($g_m$) and effective carrier mobility ($\mu_{eff}$) are two key parameters to characterize the electrical properties of graphene FET devices.[27] The transconductance is defined by $g_m = dI_D/dV_G$ where $I_D$ is the drain current and $V_G$ the gate voltage. By differentiating the $I_D - V_{BG}$ curve, $g_m$ is obtained with $V_{DS}$ = 10 mV where $V_{DS}$ is the drain-source bias. As shown in Fig. 3a, $|g_m|$ is significantly improved in GFETs on h-BN (up to 85 nS) as compared with that on SiO$_2$ (< 10 nS) for two short-channel (2.2 µm and 3 µm) GFETs.

Fig. 3b shows the effective mobility, given by $\mu_{eff} = g_d L / W Q_n$ in which the output conductance ($g_d = dI_D/dV_{DS}$) is obtained by differentiating $I_D$ with respect to $V_{DS}$ within a range of applied back gate voltage. $Q_n = C_{ox}(V_{BG} - V_{Dirac})$ is the carrier density in graphene with the oxide capacitance $C_{ox} = \varepsilon_r \varepsilon_0 / t_{ox}$ where $V_{Dirac}$ is the gate voltage at the Dirac point. The top and bottom lines in Fig. 3b are the effective mobility in graphene on SiO$_2$ and h-BN, respectively. Measured by Atomic Force Microscope (Veeco Dimension 3100), the actual thickness of the partially folded h-BN flake is found to be ~2.2 nm (along the A-B line in the AFM image, Fig. 3c). The thickness of h-BN underneath the GFET is negligible (as compared with that of the thermal SiO$_2$ layer) in the calculation of charge density in graphene, i.e., $Q_n = \frac{C_{SiO_2} C_{hBN}}{C_{SiO_2} + C_{hBN}} \cdot |V_{BG} - V_{Dirac}| \approx C_{SiO_2} \cdot |V_{BG} - V_{Dirac}|$. The carrier mobility in graphene on h-BN is at least four times higher than that on SiO$_2$.

For comparison, GFET was fabricated with mechanically exfoliated monolayer graphene on SiO$_2$ substrate. The layer number is identified via optical microscope and Raman spectrum. Both channel length and width of the fabricated GFET are about 2 µm. The small-signal transconductance of GFET with CVD-grown graphene on h-BN substrate is within a factor of 2 as compared with that of the exfoliated one (red line in Fig. 3a). In terms of high-frequency RF applications in which $g_m$ is a critical parameter, CVD-grown graphene on h-BN can be an excellent choice.

Fig. 4 shows that the performance (in terms of $g_m$ and $\mu_{eff}$) of GFET with CVD-grown graphene on h-BN is comparable to that of GFET with exfoliated graphene on SiO$_2$. The effective mobility is plotted at the same ($V_{BG}$ - $V_{Dirac}$). A carrier density is estimated to be $n = 2.16 \times 10^{12}$ cm$^{-2}$ where $n = C_{SiO_2} \cdot |V_{BG} - V_{Dirac}| / e$ with $C_{SiO_2} = 3.45 \times 10^{-8}$ F/cm$^2$, $V_{BG} - V_{Dirac} = -10$ and $e = 1.6 \times 10^{-19}$ C. The Coulomb scattering by the charged impurities in the interface between graphene and substrate[28] is considerably screened by h-BN, so that the transconductance of GFET on h-BN is higher than that of GFETs on SiO$_2$.

In summary, CVD-assembled graphene-channel field-effect transistors were fabricated on SiO$_2$ and h-BN substrates. h-BN and graphene were identified via their unique spectral signature in Raman characterization. The small-signal transconductance and effective carrier mobility are improved by 8.5 and 4 times, respectively, for GFET with CVD-assembled graphene on h-BN as compared with that on SiO$_2$. Compared with GFET with exfoliated graphene on SiO$_2$, $g_m$ and $\mu_{eff}$ measured from CVD graphene on h-BN exhibits comparable values. While the crystalline quality of CVD-based graphene is expected to be improved in the future, the results demonstrate the potential of employing h-BN as a viable supporting substrate material for emerging applications such as large-area carbon electronics.

**Acknowledgment.** The research was partially supported by National Science Foundation (NSF) grants (ECCS-1002228 and ECCS-1028267) and the IBM Faculty Award. The authors greatly appreciate support from Dr. T. Taniguchi and Dr. K. Watanabe at Advanced Materials Laboratory, National Institute for Materials Science (NIMS), Japan.

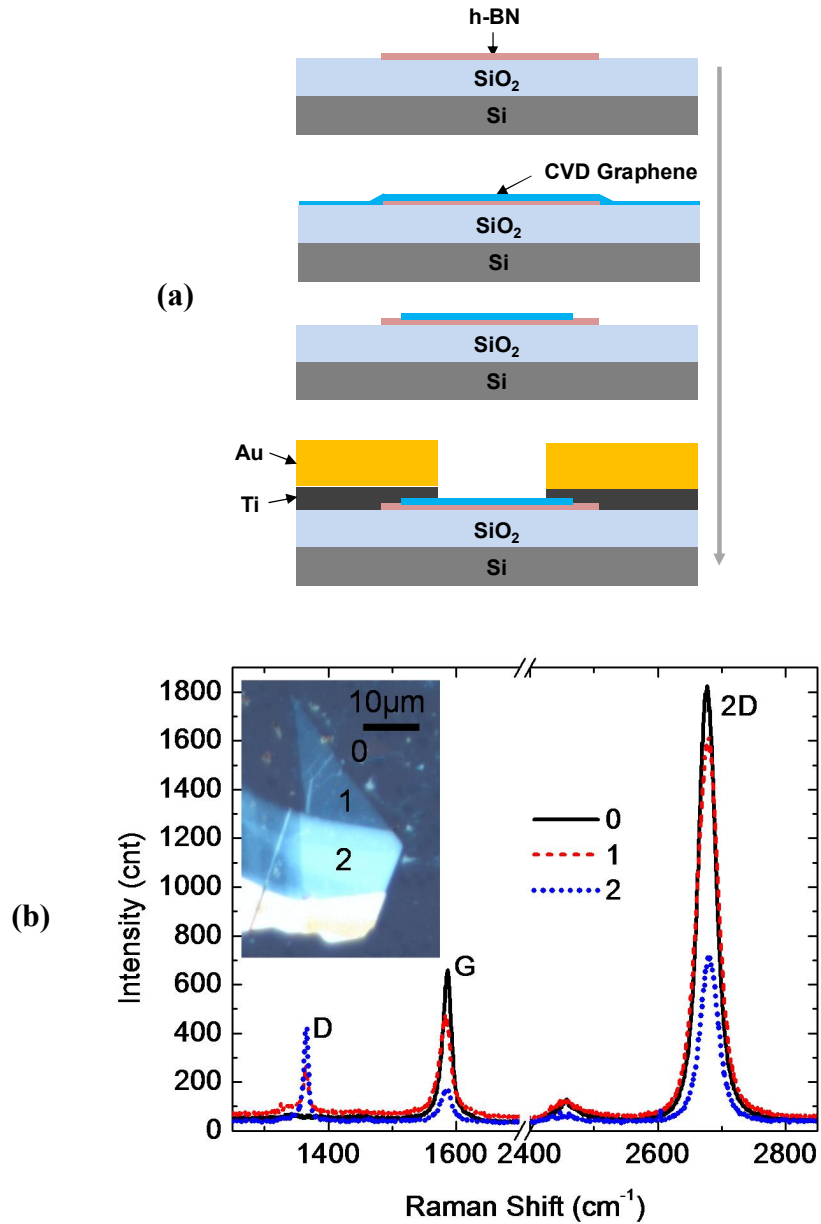

**Figure 1: (a)** Schmatic showing the fabrication process flow of GFET on h-BN. Graphene is grown by Cu-catalytic thermal CVD process. **(b)** Optical image of the prepared sample (see the inset). Position "0" represents transferred graphene with no h-BN. At positions "1" and "2", thin and thick h-BN flakes are underneath graphene, respectively. The black curve shows the Raman spectrum of graphene on $SiO_2$ at position "0". As the h-BN flake gets thicker from position "1" (red curve) to "2" (blue curve), the peaks of G- and 2D-lines shift down, while the peak of D-line increases.

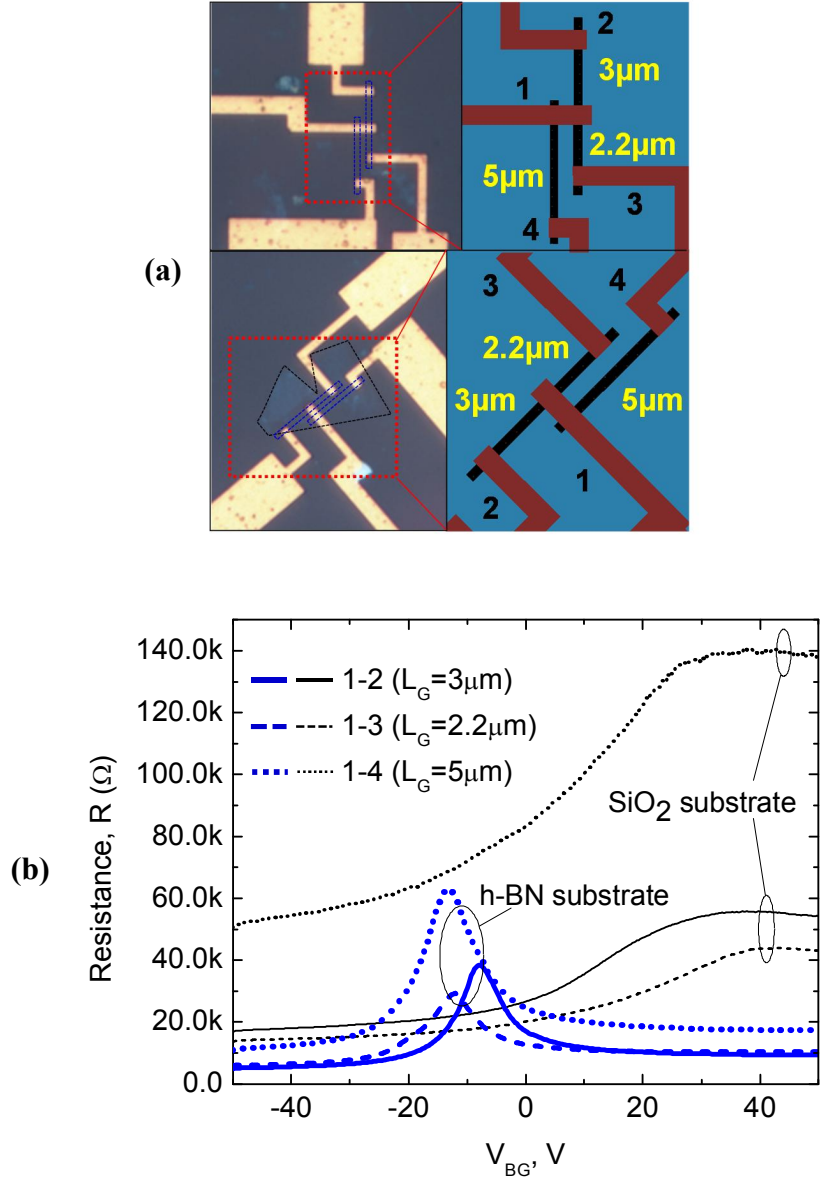

**Figure 2:** **(a)** Optical images of the fabricated GFETs with CVD-assembled graphene on SiO$_2$ (top) and on h-BN (bottom), respectively. The devices *1-2*, *1-3*, and *1-4* have different gate lengths with a shared source contact. The positions of exfoliated h-BN flake and CVD-assembled graphene are marked with black and blue dotted lines in the order. **(b)** Measured *R-V$_{BG}$* characteristics of the GFETs with CVD-assembled grapehe on SiO$_2$ and h-BN substrate, respectively.

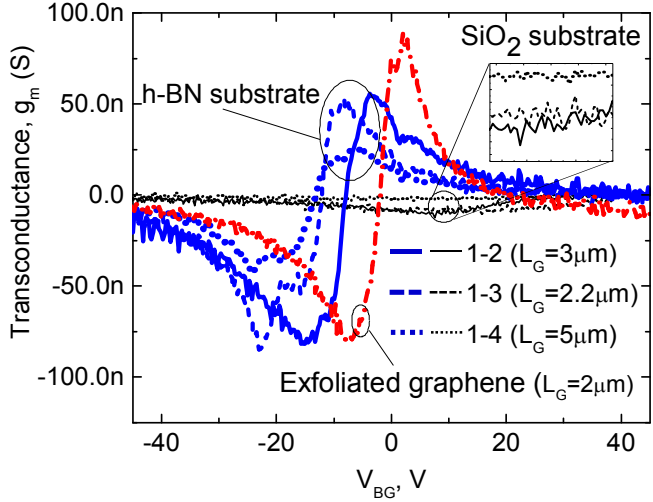
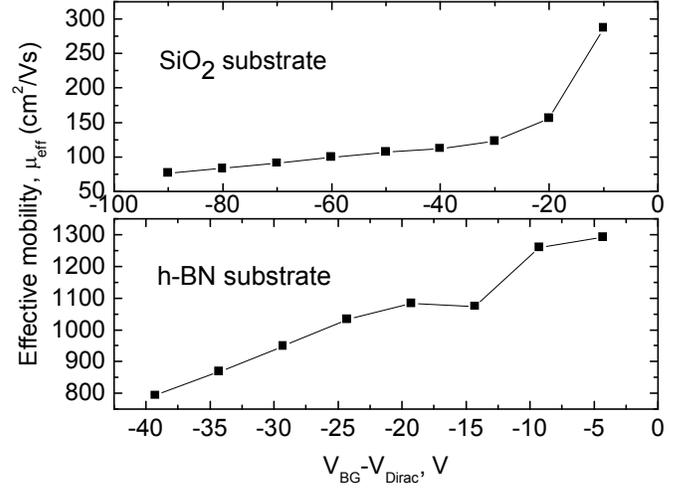

(a)

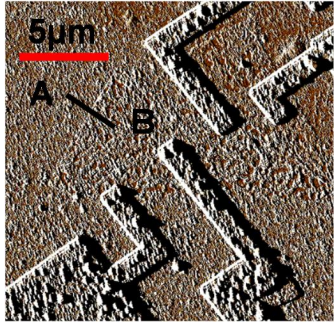
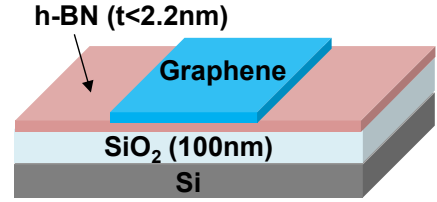

(c)

(d)

**Figure 3: (a)** Small-signal transconductances of the GFETs with CVD-assembled graphene on two different substrates. $|g_m|_{max}$ of the GFETs (gate length: 3 μm) on $SiO_2$ and h-BN are 10 nS and 85 nS (device *1-2*, solid line), respectively. For comparison, $g_m$ of the GFET with exfoliated graphene on $SiO_2$ is also plotted (red line). **(b)** Extracted carrier mobility of the GFET with CVD-assembled graphene on $SiO_2$ (top) and h-BN (bottom). The effective mobility is calculated using the output conductance obtained from $I_D - V_{DS}$ curve. **(c)** AFM image of the GFET on h-BN shows that the h-BN thickness is ~2.2 nm along the A-B line (partially folded flake). **(d)** h-BN is thinner underneath active graphene channel and its thickness is ignored in carrier density calculation.

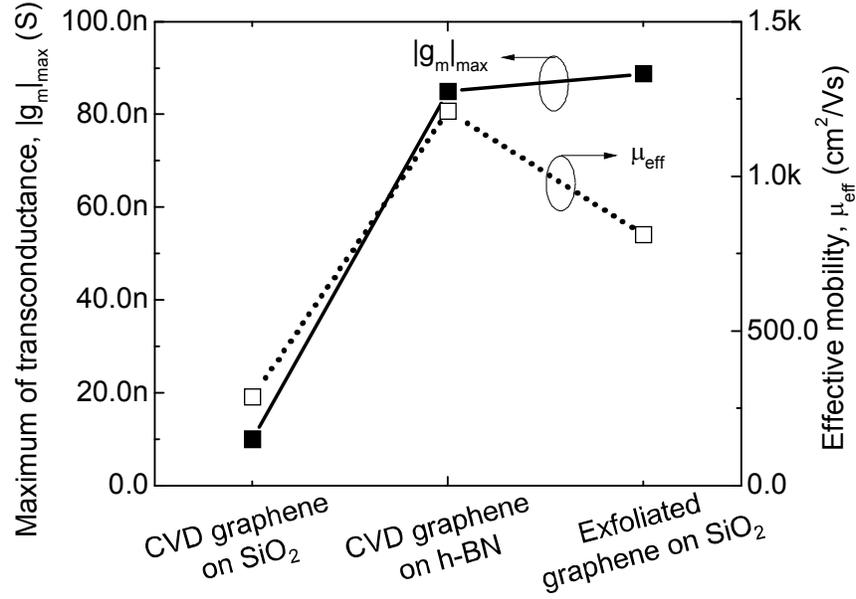

**Figure 4:** Comparison of maximum transconductance and effective carrier mobility. All three GFETs have comparable gate length (2~2.2 μm). The GFET with CVD graphene on h-BN exhibits largely improved $g_m$ and $\mu_{eff}$ over the GFET with CVD graphene on $SiO_2$, showing performance comparable to that of the GFET with exfoliated graphene on $SiO_2$. The effective mobility is measured at $V_{BG} - V_{Dirac} = -10$ V for all three devices.